\documentclass[3p,times]{elsarticle}

\usepackage{amssymb}

\journal{Nuclear Physics A}

\begin{document}

\begin{frontmatter}



\title{Thermalization and Bose-Einstein Condensation in Overpopulated Glasma}


\author[ad1]{Jean-Paul Blaizot}
\author[ad1]{Fran\c cois Gelis}
\author[ad2,ad4]{Jinfeng Liao}
\author[ad3,ad4]{Larry McLerran}
\author[ad3]{Raju Venugopalan}

\address[ad1]{Institut de Physique Th\'eorique (URA 2306 du CNRS),
  CEA/DSM/Saclay,  91191, Gif-sur-Yvette Cedex, France}
\address[ad2]{Physics Department and CEEM,
Indiana University, 2401 N Milo B. Sampson Lane, Bloomington, IN 47408, USA.}
\address[ad3]{Physics Department, Bldg. 510A, Brookhaven National Laboratory,
   Upton, NY 11973, USA}
\address[ad4]{RIKEN BNL Research Center, Bldg. 510A, Brookhaven National Laboratory, Upton, NY 11973, USA.}

\begin{abstract}
We report recent progress on understanding the thermalization of the quark-gluon plasma during the early stage in a heavy ion collision. 
The initially high overpopulation in the far-from-equilibrium gluonic matter (``Glasma'') is shown to play a crucial role. The strongly interacting nature (and thus fast evolution) naturally arises as an {\em emergent property} of this pre-equilibrium matter where the  intrinsic coupling is weak but the highly occupied
gluon states coherently amplify the scattering. A possible transient Bose-Einstein Condensate is argued to form dynamically on a rather general ground. We develop a kinetic approach for describing its evolution toward thermalization, and based on that we find approximate scaling solutions as well as numerically study the onset of condensation. 
\end{abstract}


\end{frontmatter}



\section{Introduction}

Thermalization of the quark-gluon plasma is one of the most challenging problems in current heavy ion physics.  Starting with  two colliding nuclei in a form of color glass condensate with high gluon occupation $1/\alpha_{\rm s}$ below saturation scale $Q_{\rm s}$ and following the initial impact,  a subsequent strong field evolution stage (likely with instabilities) till about the time $1/Q_{\rm s}$ is then succeeded by a far-from-equilibrium gluon-dominant matter, the Glasma. The evolution this Glasma stage toward a locally equilibrated quark-gluon plasma (QGP) is strongly indicated by phenomenology to be reached on the order of a fermi over c time. With these constraints in mind, we explore here a thermalization scenario of emergent strongly interacting matter with weak coupling albeit large aggregate of constituents \cite{Blaizot:2011xf,BLM}. We notice recent intensive discussions on this topic from a variety of approaches, see e.g. \cite{Kurkela:2011ti,Kurkela:2011ub,Epelbaum:2011pc,Gelis:2011xw,Berges:2012us,Berges:2011sb,Berges:2012ev,Berges:2012mc,Kurkela:2012hp,Schlichting:2012es,Berges:2012ks,Chiu:2012ij}. 

\section{Overpopulation as a key feature of the Glasma}

Let us start by considering the somewhat idealized problem of the evolution in a weakly coupled gluon system that is initially far from equilibrium and described by the following (Glasma-type) distribution (with coupling $\alpha_{\rm s}<<1$): 
\begin{eqnarray} \label{eq_f_Glasma}
f(p \leq Q_{\rm s}) = 1 / \alpha_{\rm s} \quad , \quad f(p>Q_{\rm s}) = 0
\end{eqnarray} 
The most salient feature of this initial gluon system, as identified in \cite{Blaizot:2011xf}, is the high overpopulation $1/\alpha_{\rm s}$, which bears a few important  consequences  by very general arguments and may hold the key of thermalization. 

The initial occupation at a value as high as $1/\alpha_{\rm s}$ coherently amplifies scattering and renders the power counting in coupling different from normal situation. For example consider the  $2\leftrightarrow 2$ gluon scattering process in the collision integral for a transport equation of gluon distribution $f(\vec p)$: while at weak coupling this process contributes at order $\sim \hat{o} (\alpha_{\rm s}^2)$ and therefore is   ``slow'' in bringing the system back to equilibrium, in the ``Glasma counting'' with $f\sim 1/ \alpha_{\rm s}$, there will be two factors from 
$f(\vec p)$ and the resulting collision term scales as $\sim \alpha_{\rm s}^2 f^2 \sim \hat{o}(1)$, despite how small the coupling $\alpha_{\rm s}$ may be. Therefore the superficially strongly interacting nature in the Glasma  (as implied by fast evolution toward equilibrium) could be an {\em emergent property} of the weakly coupled albeit highly overpopulated Glasma.    

A highly nontrivial implication of the high overpopulation is that there are so many more gluons in the Glasma than in a thermalized plasma for the same amount of energy that a Bose-Einstein Condensation(BEC) has to occur. Let us examine the overpopulation parameter, defined as a dimensionless combination of the particle number density and energy density, $ n\, \epsilon^{-3/4}$. For the Glasma distribution in Eq.(\ref{eq_f_Glasma}) one has
\begin{eqnarray}
n \sim Q_{\rm s}^3 / \alpha_{\rm s} \quad , \quad \epsilon \sim Q_{\rm s}^4 / \alpha_{\rm s} \quad , \quad (n\, \epsilon^{-3/4})_G \sim 1/  \alpha_{\rm s}^{1/4} 
\end{eqnarray}
in sharp contrast to a thermal Bose gas which has $n \sim T^3  \quad , \quad \epsilon \sim T^4    \quad , \quad n\, \epsilon^{-3/4} \sim 1$. 
More precisely  a massless Bose gas has $(n\epsilon^{-3/4})_B=\frac{30^{3/4}\, \zeta(3)}{\pi^{7/2}} \approx 0.28 <<   (n\, \epsilon^{-3/4})_G$ for realistic values $\alpha_{\rm s}\leq 0.3$. Therefore  the Glasma is significantly overpopulated from the outset and all the excessive gluons with the given amount of energy will have to be absorbed into a Bose-Einstein Condensate (BEC) if  the system evolution   is dominated by elastic processes (at least over a certain time window). With this analysis on a rather general thermodynamic ground we expect the overpopulated Glasma will thermalize to a distribution like $f_{eq}(\vec p)= n_c (2\pi)^3\delta^3(\vec p) + 1/(e^{\omega_{\vec p}/T}-1)$ with eventually a condensate density  parametrically being $n_c \sim (Q_{\rm s}^3/\alpha_{\rm s}) (1-1/\alpha_{\rm s}^{1/4})$. 

One more important feature of the initial Glasma is that there is only one scale i.e. the saturation scale $Q_{\rm s}$ which divides the phase space into two regions, one with $f>>1$ and the other with $f<<1$. One may introduce two scales for characterizing a general distribution: a soft scale $\Lambda_{\rm s}$ below which the occupation reaches $f (p<\Lambda_{\rm s}) \sim 1/\alpha_{\rm s} >>1$ and a hard cutoff scale $\Lambda$ beyond which the occupation is negligible $f (p>\Lambda) << 1$. For initial Glasma distribution one has the two scales overlapping $\Lambda_{\rm s} \sim \Lambda \sim Q_{\rm s}$. The thermalization is a process of maximizing the entropy (with the given amount of energy). The entropy density for an arbitrary distribution function is given by $s \sim \int_{\vec p} \left[(1+f)\, ln(1+f)-f \, ln (f)\right]$: this implies that with the total energy constrained, it is much more beneficial to have as wide as possible a phase space region with $f\sim 1$.  By this general argument we expect the separation of the two scales $\Lambda_s$ and $\Lambda$ in the Glasma along the thermalization process, toward the situation  for a very weakly coupled thermal gas of gluons with the soft scale $\Lambda_{\rm s}^{th} \sim \alpha_{\rm s} T $ and the hard scale $\Lambda^{th} \sim T$ separated by the coupling $\alpha_{\rm s}$.

\section{Kinetic equation and scaling solution for Glasma evolution}

With the above general insights about the evolution in overpopulated Glasma, it is tempting to demonstrate these more explicitly and quantitatively. To do that we have developed a kinetic approach assuming dominance of $2 \leftrightarrow 2$ elastic process \cite{Blaizot:2011xf,BLM}.   In the small-angle approximation one can derive the following transport equation:  
\begin{eqnarray} \label{eq_transport}
 {\mathcal D}_t f(\vec p) = \xi \left(\Lambda_{\rm s}^2 \Lambda\right)\, \vec{\bigtriangledown} \cdot  \left[ \vec{\bigtriangledown}f(\vec p) + \frac{\vec{p}}{p}\, \left(\frac{\alpha_{\rm s}}{\Lambda_{\rm s}}\right) f(\vec p)[1+f(\vec p)] \right]
\end{eqnarray}
where $\xi$ is an order one constant and the two scales $\Lambda$ and $\Lambda_{\rm s}$ are introduced and defined as:
\begin{eqnarray}
\Lambda \left( {{\Lambda_{\rm s}} \over {\alpha_{\rm s}}} \right)^2  \equiv (2\pi^2)\, \int {{d^3p} \over {(2\pi)^3}}\, f\left(\vec p\right)[1+f\left(\vec p\right)]   \quad , \quad
\Lambda \left({{\Lambda_{\rm s}} \over {\alpha_{\rm s}}} \right) \equiv  (2\pi^2)  \int {{d^3p} \over {(2\pi)^3}}\, \frac{2\, f\left(\vec p\right)}{p}
\label{eq_def2}
\end{eqnarray}
We emphasize the full nonlinearity in the $f(1+f)$ terms that arise from the Bosonic nature of gluons and become extremely crucial in the highly overpopulated case. One can see  that the Glasma distribution implies $\Lambda, \Lambda_s\sim\hat{o}(1)$ and the collision term $C\sim \Lambda_s^2 \Lambda \sim \hat{o}(1)$ in coupling, again in contrast to thermal case with $C\sim \hat{o}(\alpha_s^2)$.  
 
To describe the thermalization of the Glasma and inspired by the ``dropping-out'' of coupling, we first discuss possible scaling solution for the distribution function $f(\vec p)$ in the static box case.  We assume the following scaling form characterized by the two scales that evolve in time:
\begin{eqnarray}
f(p<\Lambda)  \sim \frac{\Lambda_{\rm s}}{\alpha_{\rm s}} \,  \frac{1}{p} \quad , \quad f(p>\Lambda) \sim 0
\end{eqnarray}
With this distribution  the coupling constant entirely drops out from the transport equation (\ref{eq_transport}) and the scattering time from the collision integral on the RHS scales as $t_{\rm sca} \sim \Lambda / \Lambda_{\rm s}^2$. To determine the time evolution of $\Lambda$ and $\Lambda_s$, we need two conditions --- that the energy must be conserved and that the scattering time shall scale with the time itself, i.e.:   
\begin{eqnarray} \label{eq_sca}
t_{\rm sca} \sim \frac{\Lambda}{ \Lambda_{\rm s}^2} \sim t   \quad  , \quad \epsilon \sim \frac{\Lambda_{\rm s} \Lambda^3 }{ \alpha_{\rm s} } =  {\rm constant} 
\end{eqnarray}
The particle number also must be conserved, albeit with a possible component in the condensate: 
$n =  n_g + n_c  \sim (\Lambda_{\rm s} \Lambda^2/ \alpha_{\rm s}) + n_c  =  {\rm constant}$.  The condensate plays a vital role with little contribution to energy while unlimited capacity to accommodate excessive gluons. Finally with the two conditions we obtain the following scaling solution:
\begin{eqnarray}
\Lambda_{\rm s}  \sim Q_{\rm s} \left( \frac{t_0}{t} \right)^{3/7} \quad , \quad  \Lambda  \sim Q_{\rm s} \left( \frac{t_0}{t} \right)^{-1/7} 
\end{eqnarray}
From this solution, the gluon density $n_g$ decreases as $\sim (t_0/t)^{1/7}$, and therefore the condensate density is growing with time, $n_c \sim (Q_{\rm s}^3/\alpha_{\rm s}) [1-(t_0/t)^{1/7}]$. A parametric thermalization time could be identified by the required  $\Lambda_{\rm s} / \Lambda \sim \alpha_{\rm s}$: 
\begin{eqnarray}
t_{\rm th} \sim \frac{1}{Q_{\rm s}} \,  \left( \frac{1}{\alpha_{\rm s}} \right)^{7/4} 
\end{eqnarray} 
At the same time scale the overpopulation parameter $n\epsilon^{-3/4}$ also reduces to be of order one.

What would change if one considers the more realistic Glasma with boost-invariant  longitudinal expansion? 
First of all the conservation laws will be manifest differently:  the total number density will decrease as $n\sim n_0 t_0/t$, while the time-dependence of energy density depends upon the momentum space anisotropy $ \epsilon \sim \epsilon_0 (t_0/t)^{1+\delta}$  for a fixed anisotropy  
 $\delta \equiv 	P_L/\epsilon$ (with $P_L$ the longitudinal pressure). Along similar line of analysis as before with the new condition of energy evolution  we obtain the following scaling solution in the expanding case:
$
\Lambda_{\rm s}  \sim Q_{\rm s} \left( t_0/t \right)^{(4+\delta)/7} \, , \,  \Lambda  \sim Q_{\rm s} \left( t_0/t \right)^{(1+2\delta)/7} 
$. 
With this solution, we see the gluon number density $n_g \sim (Q_{\rm s}^3/\alpha_{\rm s}) (t_0/t)^{(6+5\delta)/7}$, and therefore with any $\delta > 1/5$ the gluon density would drop faster than $\sim t_0/t$ and there will be formation of the condensate, i.e. 
$n_c \sim (Q_{\rm s}^3/\alpha_{\rm s}) (t_0/t) [1-(t_0/t)^{(5\delta-1)/7}]$. Similarly a thermalization time scale can be identified through the separation of scales to be: $t_{\rm th} \sim \frac{1}{Q_{\rm s}} \,  \left( \frac{1}{\alpha_{\rm s}} \right)^{7/(3-\delta)}$. Of course the possibility of maintaining a fixed anisotropy during the Glasma evolution is not obvious but quite plausible due to the large  scattering rate $\sim \Lambda_{\rm s}^2/\Lambda \sim 1/t $ that is capable of competing with the $\sim 1/t$ expansion rate and may reach a dynamical balance: see \cite{Blaizot:2011xf,Liao:2012qk} for detailed discussions.

\begin{figure*}[h]
	\begin{center}
	 \begin{minipage}[b]{6 cm}
  	\includegraphics[width=7cm,height=9.2cm]{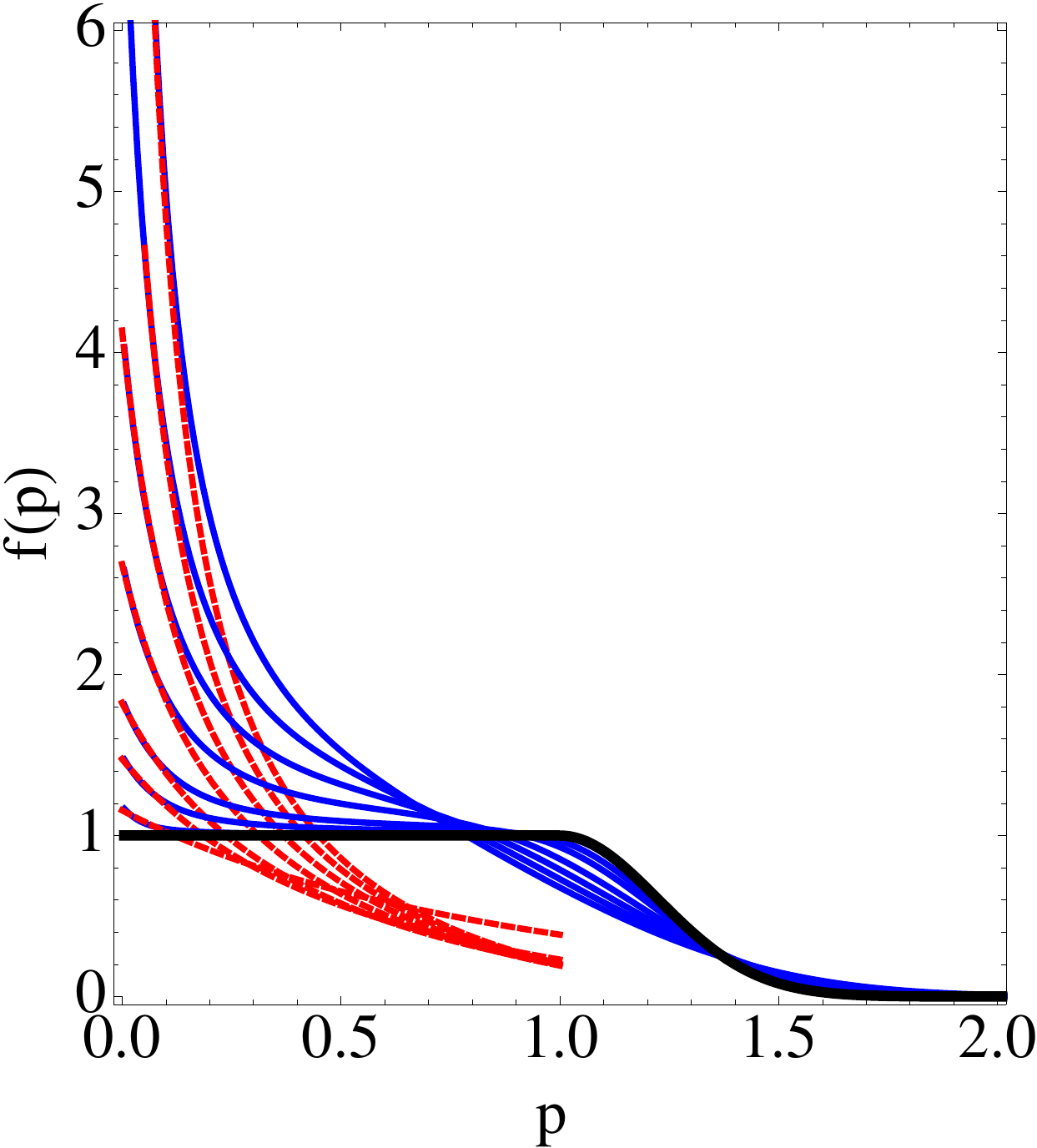}
  \end{minipage} \hspace{1.5cm}
  \begin{minipage}[b]{6 cm}
\includegraphics[width=6cm]{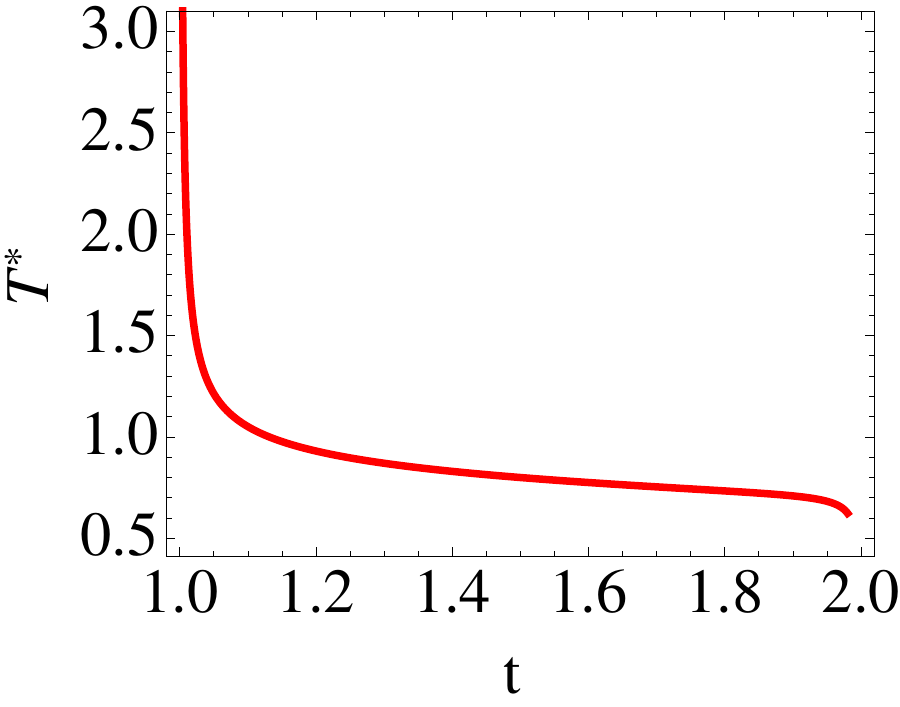}  \\
	\includegraphics[width=6cm]{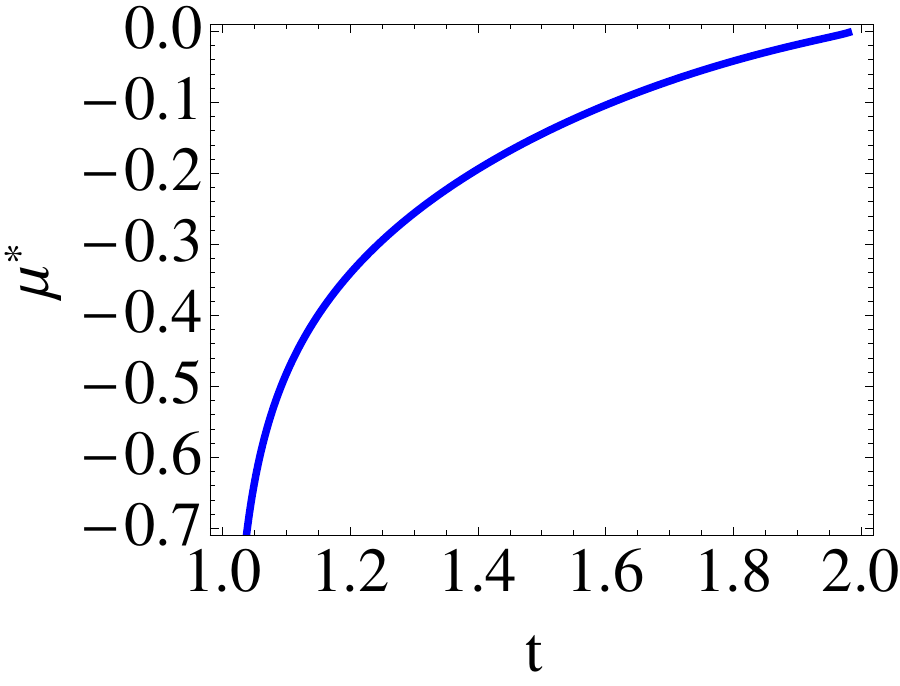} 
  \end{minipage}
		\caption{(color online)  The time evolution starting with overpopulated initial distribution toward onset of BEC: the left panel shows the distribution functions at varied time moments (as blue curves) with red dashed lines showing fitting curves $f^*=T^*/(p-\mu^*)-1/2$ for the small $p$ region; the right panel shows the time dependence of so-extracted     ``local temperature'' $T^*$ (upper) and ``local chemical potential'' $\mu^*$ (lower).}
		\label{fig_1}
			\end{center}
\end{figure*}

\section{Kinetic evolution toward the onset of condensation}

Of particular significance is to understand how dynamically the condensation occurs starting from an overpopulated initial condition. With the derived equation (\ref{eq_transport}), such question could be answered by numerically solving it. In the case without initial overpopulation, one indeed can show that the system described by this equation evolves all the way to a thermal Bose-Einstein distribution which is the proper fixed point of the collision term. In the overpopulated case, there is however the complication of the condensate formation. As is well known in atomic BEC literature, one has to separately describe the evolution prior to the onset of condensation (with this equation) and the evolution afterwards (with a coupled set of two equations explicitly for condensate and regular distribution). Efforts have been made in deriving these equations for the coevolution of a condensate and regular distribution \cite{BLM}, but here let us focus on the  pre-BEC stage  and investigate how the system approaches the onset of condensation. This stage is solely described by the Eq.(\ref{eq_transport}) and we have numerically solved it for both the static box and the expanding cases. 

With detailed results to be reported in \cite{BLM}, let us highlight the main observation. If the system starts with high initial overpopulation,  a particle flux in momentum space toward the infrared region will quickly develop and pile up particles there. The high occupation number at IR (and thus very fast scattering rate) leads to an almost instantaneous local ``equilibrium'' near the origin $\vec p=0$. This local ``equilibrium'' takes the form:  $f(\vec p\to 0) \to  T^*/(p-\mu^*)-1/2$
 with some parameters $T^*$ and $\mu^*$ one may tentatively call the local ``temperature'' and ``chemical potential''. This form is by no means a coincidence --- it consists the leading terms in the small $p$ expansion of the Bose distribution which is a fixed point of the collision term. With more and more particles being piled up near the origin, the negative local ``chemical potential'' keeps reducing its absolute value and approaches zero i.e. $(-\mu^*) \to 0^+$. This ultimately marks the onset of the condensation. All these have been explicitly seen  in the numerical solutions and shown in Fig.\ref{fig_1}. 

We have done  extensive numerical studies for varied conditions. For the static box case, the system always reaches onset as long as the initial overpopulation parameter $n\epsilon^{-3/4}$ is greater than the thermal value, $n\epsilon^{-3/4}>0.28$, despite any shape of the initial distribution (e.g. Glasma versus Guassian) or any initial anisotropy. 
  The evolution toward onset persists in the expanding case provided enough overpopulation: if starting with isotropic distribution, the critical initial overpopulation is shifted mildly to be about $n\epsilon^{-3/4}>0.40$. 
  In the expanding case with initial anisotropy: more initial longitudinal pressure $P_L>\epsilon/3$ will shift the critical overpopulation to be smaller, while less initial longitudinal pressure $P_L<\epsilon/3$ will shift it to be larger  yet only slightly even for large anisotropy. 
We therefore see that the link from initial overpopulation to the onset of condensation, in the present kinetic evolution, is a very robust one. 

Strong evidences for the formation of Bose condensate have been reported for similar thermalization problem in the classical-statistical lattice simulation of scalar field theory \cite{Epelbaum:2011pc,Gelis:2011xw,Berges:2012us}. The case for non-Abelian gauge theory is more complicated and still under investigation \cite{Berges:2011sb,Berges:2012ev,Berges:2012mc,Kurkela:2012hp,Schlichting:2012es}. In the kinetic approach, the role of inelastic processes needs further clarification though there are good arguments for the robustness of condensate even with their presence \cite{Blaizot:2011xf,Liao:2012qk}. We expect many exciting progresses yet to come along this line toward understanding the thermalization.





\bibliographystyle{elsarticle-num}







\end{document}